\documentclass{jetpl}

\usepackage{bm}
\usepackage{url}
\usepackage{tikz}

\twocolumn

\lat

\title{Wave function asymptotics for scattering of three particles with Coulomb interaction}

\rtitle{Asymptotics of the wave function for the scattering of three particles
	\ldots}

\sodtitle{Wave function asymptotics for scattering of three-particles with Coulomb interaction
}

\author{S.\,L. Yakovlev\/\thanks{e-mail: s.yakovlev@spbu.ru}
}

\rauthor{S.\,L. Yakovlev
}

\sodauthor{S.\,L. Yakovlev
}

\address{
	St Petersburg State University, St Petersburg, 199034, Russian Federation
}

\dates{27-05-2022}{*}

\abstract{The coordinate asymptotics of the wave function for the problem of scattering of three particles with Coulomb interaction is constructed. Representation of hyperspherical functions 
	is used to reduce the Schrödinger equation to a system of partial wave one-dimensional equations. Asymptotic solutions of this system are constructed by direct asymptotic methods.
}


\begin{document}
	
	\maketitle

\section{Introduction}

The coordinate asymptotics of the wave function for the scattering problem plays an decisive role in the description of particle collisions, since it is used to determine the amplitudes of colliding  processes and the corresponding cross sections. For two-particle scattering, the asymptotic form of the wave function as a superposition of incoming and outgoing spherical waves is fairly obvious. However, for three-particle collisions with free particles in the initial state, the form of the wave function asymptotics has not been fully established up to the present time \cite{MF-book}. 
Although, the asymptotics has been studied quite fully for the case of short-range interactions between particles \cite{MerkTMF-1971}, \cite{YSL-TMF-2016}, the greatest difficulty is related to the construction of the asymptotics of the wave function for the scattering of three particles with the Coulomb interaction \cite{MerkTMF-1977}. 
Numerous  attempts to go beyond the representations for leading terms of the wave function asymptotics from \cite{MerkTMF-1977}, made in \cite{Klar}, \cite{Alt} (and in a number of subsequent works by the authors and their followers), can hardly be considered successful due to insufficient completeness of these papers approaches. We should also mention monograph \cite{Peterkop}, in which a  formalism of hyperspherical coordinates was systematically applied to construct approximations for 
solution of the problem of ionization of atoms by  electron impact.

In a recent paper by the author \cite{YSL-TMF-2021} a new rigorous approach for constructing asymptotic solutions for the three-particle Schrödinger equation with short-range interaction potentials is proposed based on the weak asymptotics formalism and Faddeev integral equations.
This approach made it possible to obtain the correct asymptotics of the partial components of the wave function in the representation of hyperspherical harmonics and, as a consequence, to obtain the correct asymptotic boundary conditions for the Schrödinger equation in the representation of hyperspherical functions. At the same time, this result opens the way for constructing the asymptotics of the wave function using the asymptotics of the partial components, which are the solution to the system of one-dimensional partial wave equations. The latter significantly simplifies the formalism, since it reduces the problem of constructing the asymptotics of the solution of the multidimensional Schrödinger equation to a well-controlled procedure for finding the asymptotics of solutions to the system of one-dimensional differential equations for partial components. 
Although methods for expanding solutions of the scattering problem for three particles with short-range interactions in terms of the basis of hyperspherical functions were developed in a number of works, however, sufficient rigor of the results was not achieved, and the Coulomb case remained unexplored.
 In this regard, we mention here the work \cite{KesRos}, as well as references therein. 
In this paper, a new approach from \cite{YSL-TMF-2021} is implemented to construct the wave function asymptotics for the three-particle scattering problem with Coulomb interaction based on rigorous asymptotic methods.

\section{The Schr\"odinger equation in the representation of hyperspherical functions
}
The Hamiltonian of a system of three particles in reduced (normalized to the corresponding mass factors \cite{MF-book}) Jacobi coordinates has the form
\begin{equation*}
H=-\Delta_{\bm{x}_\beta} -\Delta_{\bm{y}_\beta} +\sum_{\gamma=1}^{3}V_\gamma(\bm{x} _\gamma).
\label{H}
\end{equation*}
Three pairs of Jacobi vectors $\bm{x}_\beta,\, \bm{y}_\beta$ c $\beta=1,2,3$, connected to each other by orthogonal transformations, form a configuration space
 vectors $\bm{X}=\{\bm{x}_\beta,\, \bm{y}_\beta \} \in \mathbb{R}^6$. 
 Hyperspherical coordinates consist of the hyperradius
$R=\sqrt{\bm{x}_\beta^2+\bm{y}_\beta^2}$ and angular variables.  
 Angular variables  of Jacobi vectors ${\hat x}_\beta, {\hat y}_ \beta$ and the hyperangle $\alpha_\beta= \tan^{-1} (y_\beta/x_\beta)$ can be chosen as the angular part of the hyperspherical coordinate set.   
Here and below, the modules of vectors are denoted in non-bold font, for example, $ x=|\bm{x}|$, and the symbol
such as ${\hat x}=\bm{x}/x$ denotes the unit vector in the direction of the vector $\bm{x}$. The well-known form of the six-dimensional Laplace operator
$\Delta= \Delta_{\bm{x}_\beta} +\Delta_{\bm{y}_\beta}$ in hyperspherical coordinates
$R, \alpha_\beta,{\hat x}_\beta, {\hat y}_\beta$ reads 
\begin{equation*}
\Delta= R^{-5}\partial_R R^5\partial_R - R^{-2} {\bf{K} }^2,
\label{DeltaHS}
\end{equation*}
where the squared hypermomentum  operator ${{\bf K}}^2$ acts only on the variable ${\hat X}=\{\alpha_\beta, {\hat x}_\beta, {\hat y}_\beta \}$.
For the purpose of this paper, the explicit form of this operator does not play a significant role, while the sets of its eigenfunctions and eigenvalues satisfying the equation
 \begin{equation}
 {\bf K}^2 Y_{\cal{K}}({\hat X})= \ell_k(\ell_k+1)Y_{\cal{K}}({\hat X}), 
 \label{K^2}
 \end{equation}
are the main components of the formalism of hyperspherical function expansions. In the formula (\ref{K^2})
$k=0,1,2,...$ are integers, $\ell_k=k+3/2$, and the multi-index $\cal{K}$ denotes the complete set of quantum numbers specifying eigenfunctions, for example,
${\cal K}=\{k,\ell_{x},m_x,\ell_y,m_y\}$, where $\ell_x, m_x$ are the momentum quantum number and its projections for the square of the orbital momentum operator corresponding to the vector $\bm{x}$, and similarly for the vector $\bm{y}$. In what follows, we will systematically use such multi-indices to specify the partial components of the wave function and the matrices representing the Hamiltonian and other quantities. A rather general and  invariant approach to the construction of hyperspherical functions $Y_{\cal K}$ can be found in \cite{Vil}.
Relations of orthogonality 
$$
\int \mbox{d}{\hat X} Y_{\cal K}({\hat X})Y^*_{\cal K'}({\hat X})= \delta_{{\cal K}{\cal K'}}
$$
and completeness 
$$
\sum_{\cal K}Y^*_{\cal K}({\hat X})Y_{\cal K}({\hat X}') =\delta({\hat X},{\hat X}'). 
$$
allow us to obtain the representation of hyperspherical functions for the wave function 
\begin{equation*}
\Psi(\bm{X},\bm{P})={(PR)}^{-5/2}\sum_{{\cal K},{\cal N}} \Psi_{{\cal K}{\cal N}}(R,P) Y_{\cal K}(\hat X) 
Y^*_{\cal N}(\hat P)
\label{Psi-pw}
\end{equation*}	
and for the Schr\"odinger equation 
\begin{multline}
\left(-\frac{d^2}{dR^2} + \frac{\ell_k(\ell_k+1)}{R^2}-P^2\right)\Psi_{{\cal K}{\cal N}}(R,P)=\\ 
-\sum_{\cal K'} V_{{\cal K}{\cal K'}}(R)\Psi_{{\cal K'}{\cal N}}(R,P). 
\label{CCE}
\end{multline}
Natrix elements for the potentials are given by integrals  
\begin{equation*}
V_{{\cal K}{\cal K'}}(R)= \sum_{\beta=1}^3 \int \mbox{d} {\hat X}\, 
Y^*_{\cal K}({\hat X})V_{\beta}(\bm{x}_\beta)Y_{\cal K'}({\hat X}).  
\label{V-KK'}
\end{equation*}
The vector  $\bm{P}\in \mathbb{R}^6$ is composed 
from relative momenta 
$\bm{q}_\beta,  \bm{p}_\beta$,  which determine the free dynamics of particles in the incident configuration. These  momenta are conjugate to the Jacobi vectors  
$\bm{x}_\beta, \bm{y}_\beta$. 
In this work by short-range potential we mean such a function $V(\bm{x})$, which behaves  as $V(\bm{x})\sim O(x^{-3-\rho}))$, $\rho>0$,  when $x\to \infty$.

In \cite{YSL-TMF-2021}, based on the analysis of the Faddeev equations for the case of short-range potentials, it is strictly obtained
the asymptotic behavior  as $PR\to \infty$ of the partial components $\Psi_{{\cal K}{\cal N}}(R,P)$ of the wave functions of the scattering problem 
for three particles free in the initial state. We present this result here as a basis for further generalization to the case of Coulomb interaction potentials between particles.
For the purpose of this paper, the asymptotics of the partial components from \cite{YSL-TMF-2021} can be conveniently written as	
\begin{multline}
\Psi_{{\cal K}{\cal N}}(R,P)\sim 
\frac{i^{k+1}}{\sqrt{2\pi}}
\left[e^{-i(RP-\ell_k\pi/2)}\delta_{{\cal K}{\cal N}} - \right. 
\\ \left. -e^{+i(RP-\ell_k\pi/2)}S_{{\cal K}{\cal N}}\right]. 
\label{Psi-as-Q0}
\end{multline}
Here $S_{{\cal K}{\cal N}}$ are the matrix elements of the $S$-matrix corresponding to transitions between states in which the particles are free both before and after interaction. 
It follows from (\ref{Psi-as-Q0}) that the functional form of this asymptotics is defined by the solutions of equations (\ref{CCE}) where  only the most leading terms has been kept as $PR\to \infty$
\begin{equation}
 \left(-\frac{d^2}{dR^2} -P^2\right)e^{\pm i(RP)}=0.  
 \label{EqQ}
 \end{equation}
It is important to note that despite the fact that there is no centrifugal term in the equation (\ref{EqQ}), additional phases in waves (\ref{Psi-as-Q0}) implicitly contain information about this term,
since the asymptotic equality holds
$$
h^\pm_{\ell_{k}}(PR) \sim  e^{\pm i(RP-\ell_k\pi/2)},  
$$
in which the Riccati-Hankel functions $h^\pm_{\ell_k}(PR)$  \cite{AbramStig} obey the equations with the centrifugal term 
\begin{equation*}
\left(-\frac{d^2}{dR^2} +
\frac{\ell_k(\ell_k+1)}{R^2} -P^2\right)h_{\ell_k}^\pm(R,P)=0. 
\label{EqQL}
\end{equation*}
The asymptotics (\ref{Psi-as-Q0}) can now be replaced by an equivalent one in terms of the Riccati-Hankel functions
\begin{equation}
\Psi_{{\cal K}{\cal N}}(R,P)\sim 
\frac{i^{k+1}}{\sqrt{2\pi}}  \left[ h^-_{\ell_k}(PR)\delta_{{\cal K}{\cal N}} - h^+_{\ell_k}(PR)S_{{\cal K}{\cal N}}  \right].
\label{Psi-as-QL}
\end{equation}
The function in the right hand side of (\ref{Psi-as-QL}) satisfies the equations (\ref{CCE}) at $PR\to \infty$ with an error determined only by the matrix elements of the potentials
$O(V_{{\cal K}{\cal N}}(R))\sim O(R^{-3-\rho})$, which can be much smaller than the original  one $O(R^{-2 })$ for the right hand side of (\ref{Psi-as-Q0}).
As a result, we have constructed the asymptotics of the solution of the scattering problem for the equations (\ref{CCE})
as a superposition of asymptotic solutions of these equations, given by the Riccati-Hankel functions $h^\pm_{\ell_k}$, with an error that decreases quite rapidly as $PR\to \infty$.  
\section{Construction of asymptotic solutions and asymptotics of the solution of the scattering problem for the Schrödinger equation with Coulomb potentials}
The main goal of this paper is to generalize the asymptotic representations (\ref{Psi-as-Q0}) and 
(\ref{Psi-as-QL})
for the case of Coulomb interactions between particles
\begin{equation*}
V_\beta(\bm{x}_\beta)=\frac{c_\beta}{x_\beta},
\label{V-C}
\end{equation*}
since in their  original form these asymptotic representations are not valid in the Coulomb case. The set of equations (\ref{CCE}) for the case of Coulomb interactions takes the form
\begin{multline}
\left(-\frac{d^2}{dR^2} +  \frac{\ell_k(\ell_k+1)}{R^2}-P^2\right)\Psi_{{\cal K}{\cal N}}(R,P)+\\ 
+ \sum_{{\cal K'}}\frac{C_{{\cal K}{\cal K'}}} {R}\Psi_{{\cal K'}{\cal N}}(R,P)=0 ,  
\label{CCE-Coulomb}
\end{multline}
where the matrix elements of the charge matrix $C_{{\cal K}{\cal K'}}$ are given by integrals 
\begin{equation}
C_{{\cal K}{\cal K'}}=  
\sum_{\beta=1}^3 c_\beta  \int \mbox{d} {\hat X}\,  
 \frac{Y^*_{\cal K}({\hat X})Y_{\cal K'}({\hat X})}{\cos \alpha_\beta} .  
\label{C-KK}
\end{equation}
Since $\mbox{d}{\hat X}=(\sin\alpha_\beta)^2(\cos\alpha_\beta)^2 d\alpha_\beta d{\hat x}_\beta d{\hat y}_\beta$ for each $\beta$, 
the integrands in (\ref{C-KK}) are not singular.  
For what follows, it is useful to pass to matrix notation. Let us introduce the matrices $\mathbb{L}$, $\mathbb{C}$ and the matrix function $\mathbb{F}(R,P)$ with matrix elements of the form	 
\begin{equation*}
[\mathbb{L}]_{{\cal K}{\cal K'}}=\ell_k(\ell_k+1)\delta_{{\cal K}{\cal K'}},\ \ [\mathbb{C}]_{{\cal K}{\cal K'}}=C_{\cal {K}{\cal K'}}, \
\end{equation*}
and 
\begin{equation*}
[\mathbb{F}]_{{\cal K}{\cal N}}(R,P)=\Psi_{{\cal K}{\cal N}}(R,P).
\end{equation*}
With these matrices let us 
rewrite the system of equations (\ref{CCE-Coulomb}) as
\begin{equation}
\left( -\frac{d^2}{dR^2}-P^2 +\frac{1}{R}\mathbb{C}+\frac{1}{R^2}\mathbb{L}\right)\mathbb{F}(R,P)=0.
\label{CCE-Matr}
\end{equation}
We construct the asymptotic solutions of the equation (\ref{CCE-Matr}), which are necessary for the scattering problem, in two stages.
At the first stage, we use the fact that, as shown in the previous section,
the leading terms of the asymptotic solutions must be determined by the solutions of the equations (\ref{CCE-Matr}), in which terms of order $O(R^{-2})$ and less are neglected  
\begin{equation}
\left( -\frac{d^2}{dR^2}-P^2 +\frac{1}{R}\mathbb{C}\right)\mathbb{F}_0(R,P)=0.
\label{CCE-Matr-0}
\end{equation}
Explicit solutions of this equation of the type of incoming and outgoing  waves are constructed using the charge matrix $\mathbb{C}$ diagonalization procedure 
\begin{equation*}
\mathbb{V}^\dagger \mathbb{C}\mathbb{V}=
\mathbb{D}_0, 
\end{equation*} 
where $\mathbb{D}_0$ is a diagonal matrix with matrix elements of the form 
\begin{equation*}
[\mathbb{D}_0]_{\cal KK'}=\delta_{\cal KK'}d_{\cal K}. 
\label{D_0}
\end{equation*}
The sign $\dagger$ here and below means the Hermitian conjugation of matrices.
As a result of this diagonalization, the equation (\ref{CCE-Matr-0}) is reduced to the form
\begin{equation}
\left( -\frac{d^2}{dR^2}-P^2 +\frac{1}{R}\mathbb{D}_0\right)\mathbb{\hat F}_0(R,P)=0 
\label{CCE-Matr-00}
\end{equation}
for $\mathbb{\hat F}_0(R,P)= \mathbb{V}^\dagger \mathbb{F}_0(R,P)\mathbb{V}$ with diagonal charge matrix.
Matrix elements of solutions for the equation (\ref{CCE-Matr-00}) of the type of incoming and outgoing waves  we  choose in the form
\begin{equation}
[\mathbb{\hat U}_0^\mp(PR)]_{\cal KK'}= \delta_{\cal KK'} \frac{i^{k+1}}{\sqrt{2\pi}}u_0^\mp(\eta_{\cal K},PR-\ell_{k}\pi/2). 
\label{E0}
\end{equation} 
Here $\eta_{\cal K}=d_{\cal K}/(2P)$, and the functions $u_0^\pm(\eta,z)$ are given in terms of regular $F_0$ and irregular $G_0$ Coulomb functions \cite{AbramStig}
by formulas
$$
u^{\pm}_0(\eta,z)= e^{\mp i\sigma_0(\eta)}[G_0(\eta,z)\pm i F_0(\eta,z)], 
$$
where $\sigma_0(\eta)=\arg \Gamma(1+i\eta)$, and have the following asymptotics as $|z| \to \infty$
\begin{equation*}
u^{\pm}_0(\eta,z) \sim e^{\pm i (z-\eta \ln2z)}. 
\label{u-0-as}
\end{equation*} 
Using the matrices $\mathbb{\hat U}^\pm_0$, we construct
the solution to the equation  (\ref{CCE-Matr-00})
as the following linear combination
\begin{equation*}
\mathbb{\hat F}_0(R,P) =  \mathbb{\hat U}^-_0(R,P) - 
\mathbb{\hat U}^+_0(R,P)  \mathbb{\hat S}_c,  
\label{F-as-0}
\end{equation*}
which gives the result for the solution of  the non-diagonalized equation (\ref{CCE-Matr-0}) 
\begin{equation}
\mathbb{F}_0(R,P) =  \mathbb{V}\mathbb{\hat U}^-_0(R,P)\mathbb{V}^\dagger - 
\mathbb{V}\mathbb{\hat U}^+_0(R,P) \mathbb{V}^\dagger   \mathbb{V}\mathbb{\hat S}_c\mathbb{V}^\dagger. 
\label{F-as-0}
\end{equation}
At the same time, since the right hand side of (\ref{F-as-0}) is a direct analog of the formula (\ref{Psi-as-Q0}) for the Coulomb case, then $\mathbb{F}_0$ gives
the leading term of the asymptotics of the solution of the scattering problem for the complete equation (\ref{CCE-Matr})
\begin{equation}
\mathbb{F}(R,P) \sim \mathbb{F}_0(R,P).
\label{F-as-F-0}
\end{equation} 
As in the case of short-range potentials,  in these formulas the quantity
$\mathbb{S}_c= \mathbb{V}\mathbb{\hat S}_c\mathbb{V}^\dagger$ has the meaning of the $S$-matrix corresponding to transitions between configurations
 in which all three particles are in states of the continuous spectrum. It should be emphasized that
due to the fact that the Coulomb functions $u_0^\pm (\eta,z) $ have the property $u_0^\pm (0,z)=e^{\pm i z }$
the formula (\ref{F-as-F-0}) becomes identical to (\ref{Psi-as-Q0}) for $\mathbb{C}=0$. As in the case of short-range potentials, the right side of (\ref{F-as-F-0})
satisfies the full equation
(\ref{CCE-Matr}) with error $O(R^{-2})$.
Thus, we have constructed the leading terms of the asymptotics of the partial components of the wave function for a system of three particles with Coulomb interaction as a direct generalization of the asymptotics
(\ref{Psi-as-Q0}).

Let us turn  to the second stage of constructing asymptotic solutions of the equation (\ref{CCE-Matr}), where we take into account the centrifugal term.
The matrices $\mathbb{C}$ and
$\mathbb{L}$ do not commute in the general case. Therefore,  the desired solutions cannot be constructed by simply replacing the functions $u^\pm_0$ in formulas like (\ref{F-as-0}) by the functions
$$
u^\pm_{\ell}(\eta,z)=e^{\mp i \sigma_\ell (\eta)}[G_{\ell}(\eta,z) \pm i F_\ell(\eta,z)],
$$
where $F_\ell$ and $G_\ell$ are the Coulomb functions corresponding to the momentum  $\ell$, and $\sigma_\ell(\eta)=\arg\Gamma(\ell +1 + i\eta)$,
just as it was done with help of Riccati-Hankel functions in the previous section for the case of short-range potentials. 
For solving the problem we use a method, which is similar to the  adiabatic expansions. We diagonalize the $R$-dependent matrix
$\mathbb{C}+ \frac{1}{R} \mathbb{L} $ 
\begin{equation*}
\mathbb{W}^\dagger(R) \left[ \mathbb{C}+ \frac{1}{R} \mathbb{L}\right] \mathbb{W}(R)= \mathbb{D}(R). 
\label{C-diag}
\end{equation*}  
Here, $\mathbb{W}(R)$ is unitary and $\mathbb{D}(R)$  is diagonal. 
Representing the solution to (\ref{CCE-Matr}) in the form 
\begin{equation}
\mathbb{F}(R,P) = \mathbb{W}(R){ \mathbb{\hat F}}(R,P)\mathbb{W}^\dagger (R), 
\label{F-hat}
\end{equation}
we obtain the following equation for $\mathbb{\hat F}(R,P)$ 
\begin{multline}
\left(-\frac{d^2}{dR^2} + \frac{1}{R}  \mathbb{D}(R)       -P^2\right) \mathbb{\hat F}(R,P)= \\
 2\mathbb{W}^\dagger(R)\frac{d\mathbb{W}(R)}{dR} \frac{d\mathbb{\hat F}(R,P)}{dR} +
 \mathbb{W}^\dagger(R)\frac{d^2\mathbb{W}(R)}{dR^2}\mathbb{\hat F}(R,P) .
\label{F-hat-eq}
\end{multline}
Since the solution of the equation (\ref{F-hat-eq}) is required for $R\to \infty$,  it suffices to construct the matrices $\mathbb{W}(R)$ and $\mathbb{D}(R)$ for large values of $R$ as the following expansions in inverse powers of $R$
\begin{eqnarray} 
\mathbb{D}(R)& = &   \mathbb{D}^{(0)} + \frac{1}{R} \mathbb{D}^{(1)}  +O(R^{-2}) , \nonumber \\  
 \mathbb{W}(R) & = &   \mathbb{W}^{(0)} + \frac{1}{R}\mathbb{W}^{(1)} +O(R^{-2}).
 \label{W} 
\end{eqnarray}
For matrices $\mathbb{D}^{(j)}$, $\mathbb{W}^{(j)}$  independent on $R$ we obtain the following 
standard perturbation theory equations
\begin{eqnarray}
\label{CW-0}
&\mathbb{C}\mathbb{W}^{(0)}&=\mathbb{W}^{(0)}\mathbb{D}^{(0)},  \\
&\mathbb{C}\mathbb{W}^{(1)}& - \mathbb{W}^{(1)}\mathbb{D}^{(0)} =  \mathbb{W}^{(0)} \mathbb{D}^{(1)}- \mathbb{L}\mathbb{W}^{(0)}.
\label{CW-1}
\end{eqnarray}
From (\ref{CW-0}) it follows that the matrix $\mathbb{W}^{(0)}$ diagonalizes $\mathbb{C }$ and hence $\mathbb{W}^{(0) }=\mathbb{V}$ and $\mathbb{D}^{(0)}=\mathbb{D}_0$. 
For matrix elements of $\mathbb{D}^{(0)}$ it results to  
$$
[\mathbb{D}^{(0)}]_{\cal KK'}=\delta_{\cal KK'}d_{\cal K}. 
$$
The solvability condition for the equation (\ref{CW-1}) in the standard way allows us to find the matrix $\mathbb{D}^{(1)}$ in the form 
\begin{equation*}
[\mathbb{D}^{(1)}]_{{\cal KK' }} = \delta_{\cal KK'} d^{(1)}_{\cal K} , \ \   d^{(1)}_{\cal K} =  \left[ \mathbb{V}^\dagger \mathbb{L} \mathbb{V}\right]_{\cal KK}.   
\label{D1}
\end{equation*}
Note that since the matrix $\mathbb{L}$ is non-negative, the diagonal elements of the matrix $\mathbb{D}^{(1)}$ satisfy the inequality
\begin{equation*}
d^{(1)}_{\cal K} \ge 0.
\label{d>0}
\end{equation*}
 The matrix elements of the solution of the equation (\ref{CW-1}) are given by the formulas
\begin{equation*}
[\mathbb{W}^{(1)}]_{\cal KK'}= -\sum_{{\cal N}\ne{\cal K'}}[\mathbb{V}]_{\cal KN} \frac{[{\mathbb{V}^\dagger\mathbb{L}\mathbb{V}]_{\cal NK'} }} {d_{\cal N}-d_{\cal K'}} . 
\label{W1}
\end{equation*}
Evaluating derivatives of  $\mathbb{W}(R)$ with the help of (\ref{W}), we arrive at expressions 
\begin{eqnarray}
\mathbb{W}^\dagger(R)\frac{d}{dR}\mathbb{W}(R)= -\frac{1}{R^2}\mathbb{V}^\dagger\mathbb{W}^{(1)} + O(R^{-3}), \nonumber \\
\mathbb{W}^\dagger(R)\frac{d^2}{dR^2}\mathbb{W}(R)= \frac{2}{R^3}\mathbb{V}^\dagger\mathbb{W}^{(1)} +O(R^{-4}).\nonumber
\end{eqnarray} 
Using these expressions, we transform the equation (\ref{F-hat-eq}) to the form  
\begin{multline}
\left(-\frac{d^2}{dR^2} + \frac{1}{R}\mathbb{D}^{(0)} + \frac{1}{R^2}\mathbb{D}^{(1)}   - P^2\right) \mathbb{\hat F}(R,P)+ \\
+\frac{2}{R^2}\mathbb{V}^{\dagger}\mathbb{W}^{(1)} \frac{d\mathbb{\hat F}(R,P)}{dR}   = O(R^{-3}), 
\label{F-hat-eq-02}
\end{multline}
where $O(R^{-3})$ accumulates  terms decreasing as $1/R^{-3}$ and faster when $R\to \infty$. We emphasize that the only non diagonal in (\ref{F-hat-eq-02}) matrix $\mathbb{V}^{\dagger}\mathbb{W}^{(1)}$
has the zero diagonal by construction.
To solve the equation (\ref{F-hat-eq-02}), we use the methods developed in \cite{YSL-TMF-adiab} for equations of this type.
We first solve the diagonal system
\begin{equation*}
\left(-\frac{d^2}{dR^2} + \frac{1}{R}\mathbb{D}^{(0)} + \frac{1}{R^2}\mathbb{D}^{(1)}   - P^2\right) \mathbb{\check U}(R,P)= 0.
\label{Z}
\end{equation*}
We choose two sets of solutions of this equation of the type of incoming and outgoing waves consistent with (\ref{E0}) in the form
\begin{equation*}
{[\mathbb{\check U}^{\mp}(PR)]}_{\cal KK'}=
\delta_{\cal KK'}  \frac{i^{k+1}}{\sqrt{2\pi}}u^\mp_{L_{\cal K}}(\eta_{\cal K},PR).
\label{U}
\end{equation*}
Here $L_{\cal K}= -1/2+\sqrt{1/4+d^{(1)}_{\cal K}}$ is a non-negative solution of the equation $L_{\cal K}(L_{ \cal K}+1) = d^{(1)}_{\cal K}$, and
$u^\mp_{L_{\cal K}}(\eta_{\cal K},PR)$ stand for incoming and outgoing  Coulomb waves \cite{AbramStig, Messiah} satisfying the equation
$$
\left(-\frac{d^2}{dR^2} + \frac{d_{L_{\cal K}}}{R}+ \frac{L_{\cal K}(L_{\cal K}+1)}{R^2}  - P^2\right) u^\mp_{L_{\cal K}}(\eta_{\cal K},PR)= 0 
$$
and having the asymptotics as $R\to \infty$ 
\begin{equation*}
u^\mp_{L_{\cal K}}(\eta_{\cal K},PR) \sim e^{\mp i(PR - \eta_{\cal K}\ln(2PK)-L_{\cal K}\pi/2)} . 
\label{u_L-as}
\end{equation*}
It remains for us to take into account the off-diagonal term in the equation (\ref{F-hat-eq-02}).
In \cite{YSL-TMF-adiab} it has been shown that solutions to an equation such as (\ref{F-hat-eq-02}) can be constructed using the representation
\begin{equation}
\mathbb{\hat U}^\pm (R,P)= \mathbb{Z^\pm}(R,P)\mathbb{\check U}^\pm(R,P). 
\label{U-ZU}
\end{equation}
For amplitudes $\mathbb{Z^\pm}$, after substituting this expression into (\ref{F-hat-eq-02}), one obtains the following equations 
\begin{multline}
-\frac{1}{R}[\mathbb{D}^{(0)},\mathbb{Z}^\pm] +\frac{1}{R^2} [\mathbb{D}^{(1)},\mathbb{Z}^\pm]    - \frac{d^2 \mathbb{Z}^\pm}{dR^2} 
-2\frac{d\mathbb{Z}^\pm}{dR} \frac{d\mathbb{\check U}^\pm}{dR}(\mathbb{\check U}^\pm)^{-1} \\   
 +\frac{2}{R^2} \mathbb{A}\frac{d\mathbb{Z}^\pm} {dR}
+\frac{2}{R^2}\mathbb{A}\mathbb{Z}^\pm \frac{d\mathbb{\check U}^\pm}{dR}(\mathbb{\check U}^\pm)^{-1} = O(R^{-3}), \nonumber
\end{multline} 
where, for brevity, the notation $\mathbb{A}= \mathbb{V}^\dagger\mathbb{W}^{(1)}$ is introduced, and $[.,.]$ denotes the matrix commutator.
Asymptotic solution for
$\mathbb{Z}^\pm$ we construct in the form 
\begin{equation}
\mathbb{Z}^\pm(R,P)= \mathbb{Z}^{(0)\pm} + \frac{1}{R}\mathbb{Z}^{(1)\pm}  
\label{TT}
\end{equation}
and, using the asymptotisc 
\begin{equation*}
\frac{d\mathbb{\check  U}^\pm}{dR}(\mathbb{\check U}^\pm)^{-1} = \pm i P\mathbb{I} +O(R^{-1}) 
\end{equation*}
from  \cite{YSL-TMF-adiab}, where  $\mathbb{I}$ is the unite matrix,  we arrive at the equation  
\begin{multline}
\frac{1}{R}\left[\mathbb{D}^{(0)}, \left(\mathbb{Z}^{(0)\pm}+ \frac{1}{R} \mathbb{Z}^{(1)\pm}\right) \right] +\frac{1}{R^2} [\mathbb{D}^{(1)},\mathbb{Z}^{(0)\pm} ]+ \\ 
+\frac{(\pm 2iP)}{R^2}\mathbb{Z}^{(1)\pm} + \frac{(\pm 2iP)}{R^2}\mathbb{A}\mathbb{Z}^{(0)\pm}
 = O(R^{-3}), 
\label{T12}
\end{multline}
In this equation we keep explicitly only terms up to the order $O(R^{-2})$, while the terms of the orders less than $O(R^{-2})$ are accumulated in $O(R^{-3})$.   
Equating terms with equal powers of $1/R$,  we obtain the equations for $\mathbb{Z}^{(0)\pm}$ and $\mathbb{Z}^{(1)\pm}$
\begin{eqnarray}
&&[\mathbb{D}^{(0)}, \mathbb{Z}^{(0)\pm}]=0, 
\label{T-0} \\
&&(\pm 2 i P)\mathbb{Z}^{(1)\pm}  + [\mathbb{D}^{(0)}, \mathbb{Z}^{(1)\pm}] =- [\mathbb{D}^{(1)},\mathbb{Z}^{(0)\pm}] - \nonumber \\ 
&& -(\pm2iP) \mathbb{A} \mathbb{Z}^{(0)\pm} . 
\label{T-1}
\end{eqnarray}
It follows from the first equation (\ref{T-0}) that $\mathbb{Z}^{(0)\pm}$ is an arbitrary diagonal matrix.
From the second equation (\ref{T-1}) we find the matrix elements of the matrix $\mathbb{Z}^{(1)\pm}$
\begin{equation*}
[\mathbb{Z}^{(1)\pm}]_{\cal KN} = - (1-\delta_{KN})\frac{(\pm 2 i P)[\mathbb{A}]_{\cal KN}}{d_{\cal K}-d_{\cal N}\pm 2 i P} [\mathbb{Z}^{(0)\pm}]_{\cal NN}. 
\end{equation*}  
So, formulas (\ref{U-ZU}) by construction provide  us with asymptotic solutions of the equation (\ref{F-hat-eq}) with an error of $O(R^{-3})$ and thus solve the problem of constructing asymptotic solutions, taking into account all terms  up to the order $O(R^{-2})$.
It is easy to see that in the leading order for $R\to\infty$ the relation holds
\begin{equation}
\mathbb{\hat U}^\pm(R,P) \sim \mathbb{Z}^\pm_0 \mathbb{\hat U}^\pm_0(R,P)  +O(R^{-1}). 
\label{hat U-hatU-0}
\end{equation}
The latter makes it possible to eliminate the arbitrariness  in the choice of $\mathbb{Z}^\pm_0$ in the obtained solutions by setting $\mathbb{Z}^\pm_0= \mathbb{I}$.

The asymptotic boundary conditions for the scattering problem for the equation (\ref{F-hat-eq}) are again given as the following  superposition
\begin{equation*}
\mathbb{\hat F}(R,P) \sim \mathbb{\hat U}^-(R,P) - \mathbb{\hat U}^+(R,P) \mathbb{\hat S}_c.  
\end{equation*}
For the solution of the scattering problem for the original equation (\ref{CCE-Matr}), we obtain now the resulting asymptotics
\begin{multline} 
\mathbb{F}(R,P) \sim \mathbb{W}(R)\mathbb{\hat U}^-(R,P) \mathbb{W}^\dagger(R) - \\
-\mathbb{W}(R)\mathbb{\hat U}^+(R,P) \mathbb{W}^\dagger(R) \mathbb{W}(R)\mathbb{\hat S}_c \mathbb{W}^\dagger(R). 
\label{F-as}
\end{multline}
The function in the right hand  side  of (\ref{F-as}) using (\ref{W}) for $\mathbb{W}(R)$ by construction satisfies the equation (\ref{CCE-Matr}) with an error of $O(R ^{-3})$ and thus the formula (\ref{F-as}) solves the problem
constructing an asymptotics for the solution of the scattering problem for the case of Coulomb potentials, taking into account the centrifugal term.
The quantity $\mathbb{S}_c=\mathbb{W}(R)\mathbb{\hat S}_c \mathbb{W}^\dagger(R)$, as above, plays the role of the $S$-matrix, corresponding to transitions between configurations,
in which all three particles are in continuous spectrum states.
Here we emphasize the direct inheritance of this fact to the result of \cite{YSL-TMF-2021}, in which
it has been rigorously proven that all processes associated with the formation of bound states in collisions in a system of three particles with short-range interactions do not contribute to the leading  terms of the asymptotics of the partial components of the wave functions.

Taking into account the relation (\ref{hat U-hatU-0}) and the form of the leading term for $W(R)$, it is easy to establish the following asymptotic equality
\begin{multline}
\mathbb{W}(R)[\mathbb{\hat U}^-(R,P) 
-\mathbb{\hat U}^+(R,P)\mathbb{\hat S}_c] \mathbb{W}^\dagger(R) =\\  
=\mathbb{V}[\mathbb{\hat U}^-_0(R,P) - 
\mathbb{\hat U}^+_0(R,P)\mathbb{\hat S}_c]\mathbb{V}^\dagger +O(R^{-1}). 
\label{as-as}
\end{multline}
Based on (\ref{as-as}) we can assert that asymptotically as $R\to\infty$ the formulas (\ref{F-as}) and (\ref{F-as-F-0}) are equivalent in leading order 
\begin{equation*}
\mathbb{F}(R,P)= \mathbb{F}_0(R,P) +O(R^{-1}). 
\end{equation*}
 Since it was shown above that the right hand side of (\ref{F-as-F-0}) at $\mathbb{C}=0$ turns into the right hand side of the formula (\ref{Psi-as-Q0}), then the asymptotic representations 
 (\ref{F-as}) has the same property. It is also important to emphasize that the asymptotic representation (\ref{F-as}), which takes into account the centrifugal terms in the Coulomb case,
 for $\mathbb{C}=0$ turns directly into the representation (\ref{Psi-as-QL}) for the scattering problem without Coulomb interactions, which takes into account the centrifugal terms in the equation
 (\ref{CCE}), respectively. 
 
 The $S$-matrix $\mathbb{S}_c$ appeared in the resulting asymptotic representations remains indefinite quantity within the framework of solving the problem of constructing 
 asymptotic solutions. As a standard,  calculation of $S$-matrix should be done at the stage of finding the physical solution to the scattering problem over the entire range of hyperradius 
 $0\le R< \infty$, 
 with the requirement that this solution satisfies both the zero boundary condition $\mathbb{F}(0,P)=0$ and  asymptotic boundary conditions (\ref{F-as})  obtained in this work .
 
 The asymptotic representation (\ref{F-as}) completes the solution of the problem of constructing the asymptotics of the wave function of the scattering problem for a system of three particles with Coulomb interaction in the representation of hyperspherical functions.
 
 \section*{Acknowledgment}
 This research was carried out as a part of SPbU Project INI 2021  using computational resources provided by Resource Center ``Computer Center of SPbU'' (http://cc.spbu.ru)

\end{document}